\documentclass[%
 reprint,
 amsmath,amssymb,
 pra
]{revtex4-1}

\usepackage{graphicx}
\usepackage{dcolumn}%
\usepackage{bm}% bold math
\usepackage{hyperref}% add hypertext capabilities
\usepackage{amsmath}
\usepackage{braket}

\begin{document}
\title{Correlations generated from high-temperature states: nonequilibrium dynamics in the Fermi-Hubbard model}
\author{Ian G. White}
\email{igw2@rice.edu}
\author{Randall G. Hulet}
\author{Kaden R. A. Hazzard}
\affiliation{Department of Physics and Astronomy, Rice University, Houston, Texas 77005, USA}
\affiliation{Rice Center for Quantum Materials, Rice University, Houston, Texas 77005, USA}

\begin{abstract}
We study interaction quenches of the Fermi-Hubbard model initiated from various high-temperature and high-energy states, motivated by cold atom experiments, which  currently operate above the ordering temperature(s). 
We analytically calculate the dynamics for quenches from these initial states, which are often strongly-interacting, to the non-interacting limit. 
Even for high-temperature uncorrelated initial states, transient connected correlations develop. 
These correlations share many features for all considered initial states. 
We observe light-cone spreading of intertwined spin and density correlations. 
The character of these correlations is quite different from their low-temperature equilibrium counterparts: for example, the spin correlations can be ferromagnetic. 
We also show that an initially localized hole defect affects spin correlations near the hole, suppressing their magnitude and changing their sign.
\end{abstract}

\maketitle

\section{\label{sec:intro}Introduction}

The development of correlations out of equilibrium is the topic of much recent research in AMO and condensed matter systems. 
Major areas of interest include the relaxation dynamics of a system driven out of equilibrium
\cite{Kinoshita2006,PhysRevLett.98.050405,transport-dynamics-2012,Neel-decay,PhysRevA.91.043623,PhysRevLett.100.175702,PhysRevA.83.063622,PhysRevX.5.041005,2016arXiv160906640S,Eisert2015,2016arXiv160805616F,PhysRevLett.109.260402-Freericks-FH-Efield} 
and the possibility of relaxation to nonthermal steady states 
\cite{PhysRev.109.1492,Basko20061126,general-lattice-thermalization,MBL-motivation,PhysRevB.82.174411,RevModPhys.83.863,PhysRevB.89.165104,PhysRevLett.100.030602,PhysRevA.89.033616,PhysRevLett.103.056403,Choi1547,2016arXiv161003284K} 
that have unusual properties 
\cite{PhysRevLett.95.206603,MBL-motivation,PhysRevLett.111.127201,PhysRevB.88.014206,PhysRevB.90.174202,PhysRevLett.109.017202,PhysRevLett.110.067204,PhysRevLett.110.260601,PhysRevB.90.064201,Schreiber842,PhysRevLett.114.083002}. 
An emerging direction involves inducing nonequilibrium correlations at temperatures above those required for equilibrium order. This has been demonstrated in some solid state systems in the presence of continuous driving 
\cite{Nicoletti:16}.
However, the dynamics after quenches has been less studied, and numerous questions exist in all cases: What conditions are required for correlations to develop? 
What timescales are involved? 
What will the character of these correlations be?

In this paper we study quenches of the Fermi-Hubbard model from finite (and in some cases very high) initial temperatures to noninteracting final Hamiltonians. 
This is a useful complement to studies that consider dynamics from low temperature initial conditions 
\cite{PhysRevLett.100.175702,PhysRevLett.105.076401,Demler-fast-ramp-corr,Neel-decay}.  
Besides its intrinsic interest, this regime is important to ongoing experiments.
This is because despite much recent progress towards realizing low temperature equilibrium states experimentally, the regime well below the ordering temperatures (e.g. the N\'eel temperature for the antiferromagnet) remains elusive due to the very low temperatures and entropies required
\cite{MI-Hulet,AFM-corr-Hulet,Jordens2008,Greif1307,Esslinger-thermometry-MI,PhysRevLett.112.115301,PhysRevLett.115.260401,Schneider1520,Hackermueller1621,Endres2013,Boll1257,AFM-corr-Zwierlein-QGM,2016arXiv160604089C,Greif953,Parsons1253,Taie2012,PhysRevX.6.021030,PhysRevLett.116.175301,PhysRevLett.117.135301,2016arXiv160700392D,Esslinger-Review}.

We find that, even when initiated from high temperature initial states that are above the superexchange or even tunneling energy scales, such quenches generate transient particle number and spin correlations between two sites; after the quench a light cone of connected correlations between increasingly distant sites develops over time.
A wide range of initial product states exhibit qualitatively similar correlation dynamics.
In particular we calculate the dynamics for high temperature Mott insulators in one and two dimensions, a strongly interacting metal, a partially spin-polarized Mott insulator, and a perfect product state antiferromagnet.
 
The transient correlations can be qualitatively different from the correlations of the equilibrium low temperature states of the same initial Hamiltonian. 
For example, we observe the generation of ferromagnetic spin correlations from a Hamiltonian with initially repulsive on-site interactions, in contrast to the antiferromagnetic spin correlations that occur in equilibrium for the repulsive Hubbard model.

Going forward, our results will help one understand quenches with finite interactions after the quench.
On the one hand, when phenomena persist with interactions our results provide a foundation for understanding them. 
On the other hand, when interactions lead to phenomena that are absent in our results, it signals that the physics is intrinsically interacting. 
Given how surprising out-of-equilibrium dynamics can be, it is crucial to sort out which surprises result from the interactions and which arise from the inherent nonequilibrium nature of the problem (independent of interactions).

An example drives this home. 
Imagine that one found - perhaps in a strongly interacting system - that spin and density correlations were evolving dynamically with exactly the same magnitude. 
This intriguing behavior is reminiscent of ``intertwined" spin and density order in equilibrium strongly correlated systems
\cite{RevModPhys.87.457}. 
Although a natural instinct is to imagine this observation is similarly non-trivial, one of our results is to show that such dynamics occurs even for non-interacting quenches.
Thus, remarkably intricate phenomena can occur even in the non-interacting dynamics. Comparing to this important baseline allows one to assess how dramatic a given observation in a strongly-interacting system really is.

Another interesting example that we study in this context is the transport of a hole defect after a quench.
We show that as the hole propagates it affects the development of correlations around it.
Superficially, it appears that the hole is dressed with a cloud of spin correlations.
This is another example where, if this were observed in a strongly interacting system one might leap to the conclusion that the physics was highly non-trivial, but in fact the richness here appears already in the noninteracting dynamics. 

This paper is organized as follows: Section \ref{sec:dynamics} describes how we calculate the dynamics of observables quenched from initial spatial product states to non-interacting Hamiltonians.
Section \ref{sec:MI} applies the theory to calculate the connected correlations in a one-dimensional Mott insulator. 
Section \ref{sec:compare} shows that qualitatively similar phenomena persist for multiple initial conditions. 
Section \ref{sec:hole transport} describes how an initially localized hole defect modifies the dynamics of spin correlations. 
Section \ref{sec:conclusion} presents conclusions and outlook.

\section{\label{sec:dynamics}Quench dynamics in the non-interacting limit}

\begin{figure}
\includegraphics[clip=true, trim= 0mm 3mm 180mm 150mm, width= 4.00in, keepaspectratio=true]{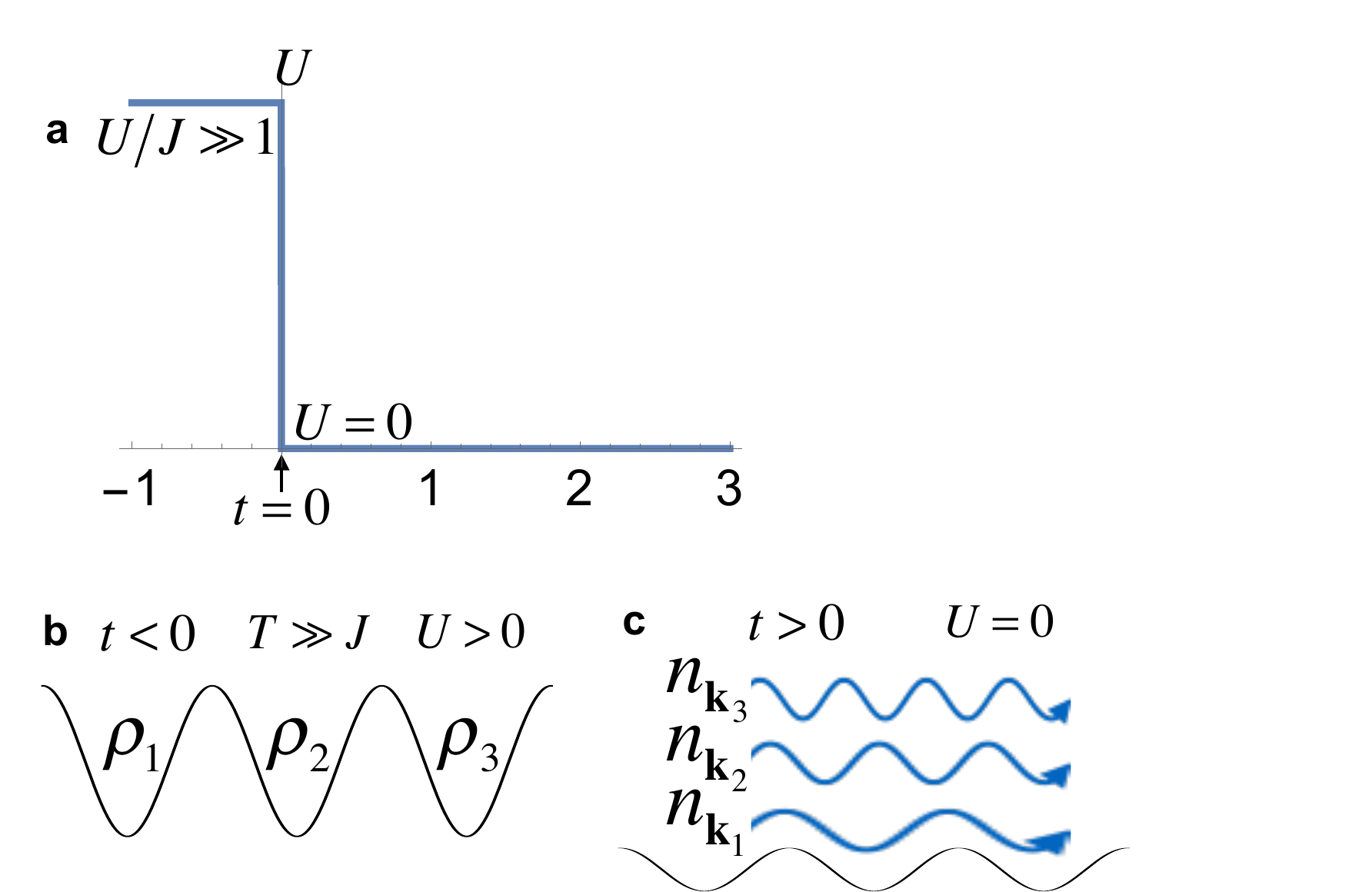}
\caption{\label{fig:quench} (a) Quench protocol for the dynamics in this paper (arbitrary units). (b) Pre-quench, the system is in a product of single-site states. An important class of states of this form that we consider arise from the $J\ll T \ll U$ equilibrium state of the Fermi-Hubbard Hamiltonian. (c) Post-quench, the system evolves in the noninteracting limit of the Fermi-Hubbard Hamiltonian, with conserved momentum occupation numbers.}
\end{figure}

We consider interaction quenches from initial product states to the noninteracting limit for ultracold fermions in an optical lattice, illustrated in Fig. \ref{fig:quench}.
The initial state is
\begin{equation}
\rho=\bigotimes_{i}\rho_{i}^{(1)}.\label{eq: density matrix}
\end{equation}
where $\rho_i^{(1)}$ is an arbitrary density matrix for site $i$ (in general a mixed state). 

The system is described by the Hubbard Hamiltonian
\begin{equation}
H=-J\underset{\left\langle ij\right\rangle ,\sigma}{\sum}c_{i\sigma}^{\dagger}c_{j\sigma}+U\underset{i}{\sum}n_{i\uparrow}n_{i\downarrow}\label{eq:H}
\end{equation}
where $\left\langle ij\right\rangle$ indicates a nearest neighbor pair of sites,  $\sigma \in \left\{ \uparrow \,, \downarrow\right\}$, $c_{i\sigma}$ is the fermionic annihilation operator at site $i$ with spin $\sigma$, and $n_{i\sigma}=c^\dagger_{i\sigma}c_{i\sigma}$ is the corresponding number operator.
This describes fermions in a deep lattice with a nearest-neighbor tunneling amplitude $J>0$ and on-site interaction energy $U$
~\cite{1998PhRvL..81.3108J}. 
Many of our initial states arise as high-temperature ($T\gg J$) equilibrium states of Eq.~\eqref{eq:H}, and the post-quench dynamics is governed by its $U=0$ limit.

Figure ~\ref{fig:quench} illustrates our quench protocol, in which the system starts in equilibrium at some value of $U$ and the interaction is turned off at $t=0$:
\begin{equation}
U(t)=U_{0}\left[1-\Theta(t)\right]
\end{equation}  
where $\Theta$ is the Heaviside step function and $\left| U_{0}\right| \gg J$.
When the temperature $T$ before the quench is large compared to $J_{\text{init}}$ (the tunneling before the quench) -- i.e. $T\gg J_{\text{init}} $ -- the initial state takes the form of Eq.~\eqref{eq: density matrix}. (We will consider a few alternative product states later.)
Experimentally, the interaction can be dynamically controlled by using a Feshbach resonance or changing the lattice depth.

We note that our calculations actually describe a variety of more general quenches of the Fermi-Hubbard model. The only required conditions are that the initial temperature $T$ satisfies $T\gg J_{\text{init}}$ and $U=0$ after the quench. So, for example, one could suddenly change both $U$ and $J$ at time $t=0$ as long as these conditions are met.

Our goal is to calculate the density and spin expectation values and two-site correlation functions for $t>0$. We define the total density operator $n_i=\sum_\sigma n_{i\sigma}$ and the spin operators  
$ \vec{S}_i = \frac{1}{2}\sum_{\alpha\beta}c_{i\alpha}^\dagger \vec{\sigma}_{\alpha\beta} c_{i\beta}^{\phantom \dagger}$, where $\vec{\sigma}$ is the vector of Pauli matrices.
We focus on these observables as the most basic correlations that characterize equilibrium systems, and because they can be measured in experiments.
Note that the correlation functions can be expressed as
\begin{eqnarray}
\langle n_{i} n_{j}\rangle &=& \sum_{\alpha\beta} \langle c^\dagger_{i\alpha}c_{i\alpha}^{\phantom \dagger}c^\dagger_{j\beta}c_{j\beta}^{\phantom \dagger} \rangle \label{eq: density correlator def} \\
\langle S_{i}^a S_{j}^b \rangle &=& \frac{1}{4}\sum_{\alpha\beta\gamma\delta}\sigma^{a}_{\alpha\beta}\sigma^{b}_{\gamma\delta}\langle c^\dagger_{i\alpha}c_{i\beta}^{\phantom \dagger}c^\dagger_{j\gamma}c_{j\delta}^{\phantom \dagger} \rangle \label{eq: spin correlator def}
\end{eqnarray}
where $a \,, b \in \left\{ x \,, y \,, z \right\}$.
Therefore we turn to calculating the dynamics of a general two-site correlation $\left\langle c_{i\alpha}^\dagger c_{i\beta}^{\phantom \dagger} c^\dagger_{j\gamma} c^{\phantom \dagger}_{j\delta}\right\rangle$, from which we can obtain the density and spin correlations.
For compactness, we define
\begin{eqnarray}
C^{nn}_{ij} &=& \braket{n_i n_j}- \braket{n_i}\braket{n_j}   \\
C^{ab}_{ij} &=& \braket{S_i^a S_j^b}- \braket{S_i^a}\braket{S_j^b} \label{eq: C notation}
\end{eqnarray}
where $a,b \in \{x,y,z\}$.
 
Because the Hamiltonian after the quench is non-interacting, one can analytically express the time-evolution of the  annihilation operator as
\begin{equation}
c_{j\alpha} (t)=\underset{l}{\sum}A_{jl}(t)c_{l\alpha}\label{eq: c operator}
\end{equation}
where $A_{jl}(t)$ 
is the propagator from site $l$ to site $j$ of a single particle on the lattice.
Eq.~\eqref{eq: c operator} follows because our Hamiltonian can be written $H=\sum_{k\alpha} {\mathcal E}_k b_{k\alpha}^\dagger b_{k\alpha}$ for some set of annihilation operators $b_{k\alpha}$. 
The time evolution of these operators is $b_{k\alpha}(t)=e^{-i{\mathcal E}_k t} b_{k\alpha}$. 
(If no time argument is provided, the operator is evaluated at $t=0$, and we set $\hbar=1$ throughout.) 
The annihilation operators $c_{j\alpha}$ can be expressed $c_{j\alpha} = \sum_k S_{j k} b_{k\sigma}$ for some $S_{j k}$. 
Conversely, $b_{k\alpha} = \sum_{j} (S^{-1})_{k j} c_{j\alpha}$. 
Hence at time $t$, $c_{j\alpha}(t) = \sum_k S_{j k} b_{k\alpha}(t) = \sum_{k} S_{j k} e^{-i {\mathcal E}_k t} b_{k\alpha} = \sum_{k l} e^{-i {\mathcal E}_k t} S_{j k} (S^{-1})_{k l} c_{l\alpha}$.

In one dimension the single particle eigenstates $k$ can be identified with quasi-momentum states in the first Brillouin zone, for which $\mathcal{E}_{k} = -2 J\cos\left(ka\right)$ and $S_{jk}=\exp{\left(ijka\right)}/\sqrt{N}$.
Taking $N \rightarrow \infty$ we see that Eq.~\eqref{eq: c operator} holds with 
\begin{equation}
A_{jl}(t)=\left(-i\right)^{\left|j-l\right|}\mathcal{J}_{\left|j-l\right|}\left(2Jt\right)\label{eq:propagator}
\end{equation}
 where $\mathcal{J}_{m}\left(z\right)$ is a Bessel function of the
first kind. 

The expectation value of the general two-site correlator that determines the density and spin correlations at time $t$ is given in terms of initial expectation values by
\begin{multline}
\hspace{-.5cm}\left\langle c_{i\alpha}^{\dagger}(t)c_{i\beta}(t)c_{j\gamma}^{\dagger}(t)c_{j\delta}(t)\right\rangle ={}\\ \hspace{1cm}\underset{p,q,r,s}{\sum}A_{ip}^{*}(t)A_{iq}(t)A_{jr}^{*}(t)A_{js}(t)\left\langle c_{p\alpha}^{\dagger}c_{q\beta}c_{r\gamma}^{\dagger}c_{s\delta}\right\rangle_0 \label{eq: time evolution of expectation value} 
\end{multline}
using Eq.~\eqref{eq: c operator}, where $\left\langle \cdots \right\rangle_0$ indicates the expectation value at time $t=0$.

We compute these initial expectation values by taking advantage of the product state nature of Eq.~\eqref{eq: density matrix}. 
In this state, expectation values of operators factor by site: 
$\left\langle P_{i}Q_{j}\right\rangle _{0}=\left\langle P_{i}\right\rangle _{0}\left\langle Q_{j}\right\rangle _{0}$
if $i \ne j$ for operators $P_i$ and $Q_j$ supported on single sites.
Then Eq.~\eqref{eq: time evolution of expectation value}
factors into a sum of three types of non-vanishing terms: (i) $p=q=r=s$, (ii) $p=q$ and $r=s$ with $p\ne r$, and (iii) and $p=s$ and $r=q$ with $p\ne r$. 
Writing the expectation in terms of these sums (and renaming summation indices) we have
\begin{widetext}
\begin{eqnarray}
\left\langle c_{i\alpha}^{\dagger}c_{i\beta}^{\phantom \dagger}c_{j\gamma}^{\dagger}c_{j\delta}^{\phantom \dagger}\right\rangle (t) & = & \sum_{p}\left|A_{ip}(t)\right|^{2}\left|A_{jp} (t )\right|^{2}\left\langle c_{p\alpha}^{\dagger}c_{p\beta}^{\phantom \dagger}c_{p\gamma}^{\dagger}c_{p\delta}^{\phantom \dagger}\right\rangle _{0}\nonumber \\
 &  & \hspace{-.8in}{}+\sum_{p\ne q}\left|A_{ip}(t)\right|^{2}\left\langle c_{p\alpha}^{\dagger}c_{p\beta}^{\phantom \dagger}\right\rangle _{0}\left|A_{jq}(t)\right|^{2}\left\langle c_{q\gamma}^{\dagger}c_{q\delta}^{\phantom \dagger}\right\rangle _{0}%
+\sum_{p\ne q}A_{ip}^{*}(t)A_{jp}(t)\left\langle c_{p\alpha}^{\dagger}c_{p\delta}^{\phantom \dagger}\right\rangle _{0}A_{iq}(t)A_{jq}^{*}(t)\left\langle c_{q\beta}^{\phantom \dagger}c_{q\gamma}^{\dagger}\right\rangle _{0}\!.\label{eq: general time evolution eq}
\end{eqnarray}
\end{widetext} 
Although the last two terms are double sums over $p$ and $q$ with $p\ne q$, the summand factors. 
The sums can be written as products of single sums because in general $\sum_{p\ne q} P_p Q_q = \sum_{p, q} P_p Q_q- \sum_p P_p Q_p$. 
Using this, and using $\left\langle c^{\phantom \dagger}_{p\alpha}c_{p\beta}^{\dagger}\right\rangle _{0}=\delta_{\alpha\beta}-\left\langle c_{p\beta}^{\dagger}c^{\phantom \dagger}_{p\alpha}\right\rangle _{0}$ to write each expectation value in a structurally similar form allows us to rewrite Eq.~\eqref{eq: general time evolution eq} as
\begin{widetext}
\begin{eqnarray}
\left\langle c_{i\alpha}^{\dagger}c_{i\beta}^{\phantom \dagger}c_{j\gamma}^{\dagger}c_{j\delta}^{\phantom \dagger}\right\rangle (t)& = & \sum_{p}\left|A_{ip}(t)\right|^{2}\left|A_{jp}(t)\right|^{2}\left[g _{\alpha\beta\gamma\delta}^{p}-f _{\alpha\beta}^{p}f _{\gamma\delta}^{p}-f _{\alpha\delta}^{p}\left(\delta_{\beta\gamma}-f _{\gamma\beta}^{p}\right)\right]\nonumber \\
 &  & \hspace{-.8in}{}+\left[\sum_{p}\left|A_{ip}(t)\right|^{2}f _{\alpha\beta}^{p}\right]\left[\sum_{q}\left|A_{jq}(t)\right|^{2}f_{\gamma\delta}^{q}\right]
+\left[\sum_{p}A_{ip}^{*}(t)A_{jp}(t)f _{\alpha\delta}^{p}\right]\left[\sum_{q}A_{iq}(t)A_{jq}^{*}(t)\left(\delta_{\beta\gamma}-f _{\gamma\beta}^{q}\right)\right]\!.\label{eq: factored time evolution eq}
\end{eqnarray}
\end{widetext}
We have defined
\begin{subequations}
\label{eq: f and g}
\begin{eqnarray}
f_{\alpha\beta}^{i} &=& \left\langle c_{i\alpha}^{\dagger}c_{i\beta}^{\phantom \dagger}\right\rangle _{0} \label{eq: f} \\
g_{\alpha\beta\gamma\delta}^{i} &=& \left\langle c_{i\alpha}^{\dagger}c_{i\beta}^{\phantom \dagger}c_{i\gamma}^{\dagger}c_{i\delta}^{\phantom \dagger}\right\rangle _{0} \label{eq: g}
\end{eqnarray}
\end{subequations}
to simplify notation.

With this rearrangement, double sums are eliminated (they factor) and only single sums remain.
In combination with Eqs.~\eqref{eq: density correlator def} and~\eqref{eq: spin correlator def} this allows the time evolution of the density-density and spin-spin correlators to be calculated efficiently.

We note that our calculations are similar to those by Gluza et al. in~\cite{PhysRevLett.117.190602} who also study noninteracting lattice fermions. 
We consider a concrete, physically relevant case and focus on interesting phenomena that occur during the transient dynamics.

\section{\label{sec:MI}Quench from  $T \gg J$ Mott insulator}

We now apply the results of Section~\ref{sec:dynamics} to a Mott insulating initial state (i.e., no spin polarization and unit filling). 
In particular we consider $T \gg J$ and (on-average) unit filling enforced by choosing the chemical potential to be $\mu=U/2$. 
Define
\begin{equation}
H_{i} = U n_{i\uparrow}n_{i\downarrow} - \mu n_{i} \label{eq:single-site H}
\end{equation}
In this limit, the density matrix is given by 
\begin{eqnarray}
\rho&=&Z^{-1}\exp\left(-\beta H\right) \nonumber \\
      &=&Z^{-1}\exp\left(-\beta\sum_i H_i + O (J/T)\right)\nonumber \\
      &\approx&Z^{-1}\bigotimes_{i}\exp\left(-\beta H_{i}\right)\label{eq: density matrix MI}
\end{eqnarray}
where $Z$ is a constant enforcing $\operatorname{Tr}\rho = 1$, $\beta=1/T$ is the inverse temperature and we set $k_B=1$ throughout.

In what follows, we will associate with any energy $A$ a dimensionless ratio $\tilde{A}=\beta A$. 
Then the expectation values in the initial state are
\begin{subequations}
\label{eq: f and g Mott}
\begin{eqnarray}
f_{\alpha\beta}^{i} &=& \frac{1}{2}\delta_{\alpha\beta} \label{eq: f Mott} \\
g_{\alpha\beta\gamma\delta}^{i} &=& \begin{cases}
	\frac{1}{2} &\alpha=\beta=\gamma=\delta\\
	\frac{1}{2}\frac{1}{1+e^{\frac{1}{2} \tilde{U}}} &\alpha=\beta\ne\gamma=\delta\\
	\frac{1}{2}\left(1-\frac{1}{1+e^{\frac{1}{2} \tilde{U}}}\right) &\alpha=\delta\ne\beta=\gamma\\
	0 &\rm{otherwise}\\
\end{cases} \label{eq: g Mott}
\end{eqnarray}
\end{subequations}

\begin{figure}
\includegraphics[clip=true, trim= 0mm 0mm 170mm 100mm, width= 4.00in, keepaspectratio=true]{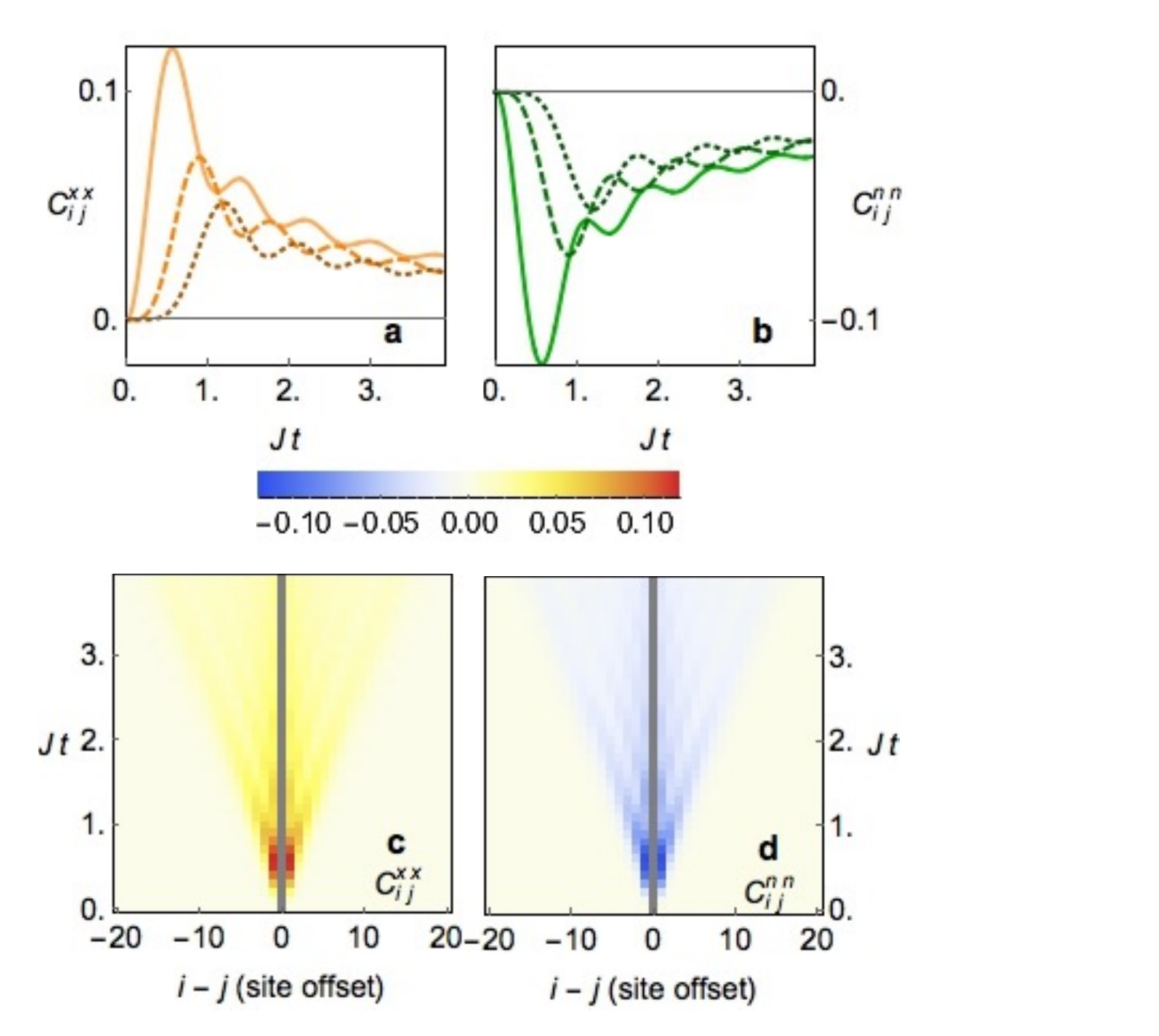}
\caption{\label{fig: MI data}Connected correlations of a $J\ll T \ll U$ 1D Mott insulator quenched to a noninteracting Hamiltonian. (a) Spin-spin correlations $C^{xx}_{ij} = \langle \sigma^x_i \sigma^x_j\rangle - \langle \sigma^x_i \rangle \langle \sigma^x_j\rangle$ and (b) density-density correlations $C^{nn}_{ij} = \langle n_i n_j\rangle - \langle n_i \rangle \langle n_j\rangle$ between sites with an offset of one (solid lines), two (dashed lines), or three (dotted lines). (c) Spin-spin and (d) density-density correlations as a function of time and site offset.}
\end{figure}

Figure \ref{fig: MI data} shows the post-quench correlation dynamics of this $T \gg J$ Mott insulating initial state obtained by Eq.~\eqref{eq: factored time evolution eq} using $f^i_{\alpha\beta}$ and $g^i_{\alpha\beta\gamma\delta}$ given in Eq.~\eqref{eq: f and g Mott}.
For a fixed distance, transient connected correlations develop after the quench.
Connected correlations of both spin [Fig. \ref{fig: MI data}(a,c)] and density [Fig. \ref{fig: MI data}(b,d)] develop as a function of time in the shape of a light cone: correlations develop inside, and at the edge of, a region in space whose size grows as $vt$ for some velocity $v$. 
We observe that connected correlations spread at a velocity $v \approx 4 J a$, which is twice the maximum group velocity of a single particle with dispersion relation $\mathcal{E}_k = -2 J \cos\left( k a \right)$. 
The correlations can spread with twice the velocity of a single particle since two lattice sites can be mutually influenced by signals from a source halfway between them. This is consistent with previous work describing the spread of correlations after a quench \cite{PhysRevLett.96.136801,Cheneau2012,PhysRevB.94.085122,1367-2630-16-5-053034,1367-2630-14-2-023008-LR-bound}. 

In contrast to the low-temperature equilibrium state, the correlations are ferromagnetic rather than antiferromagnetic. 
Furthermore, the spin and density correlations are intertwined, suggesting an emergent symmetry.
Specifically, the system develops positive spin-spin connected correlations that are independent of the spin orientation (i.e. $C^{xx}_{ij}=C^{yy}_{ij}=C^{zz}_{ij}$) and negative density-density connected correlations $C^{nn}_{ij}$ of equal magnitude.
The intertwined spin and density correlations stem from the fact that in the noninteracting dynamics, there is only one energy scale, which is set by the tunneling $J$. Thus the spin and density correlations are controlled by the same energy scale.

\begin{figure}
\includegraphics[clip=true, trim= 0mm 3mm 220mm 160mm, width= 4.00in, keepaspectratio=true]{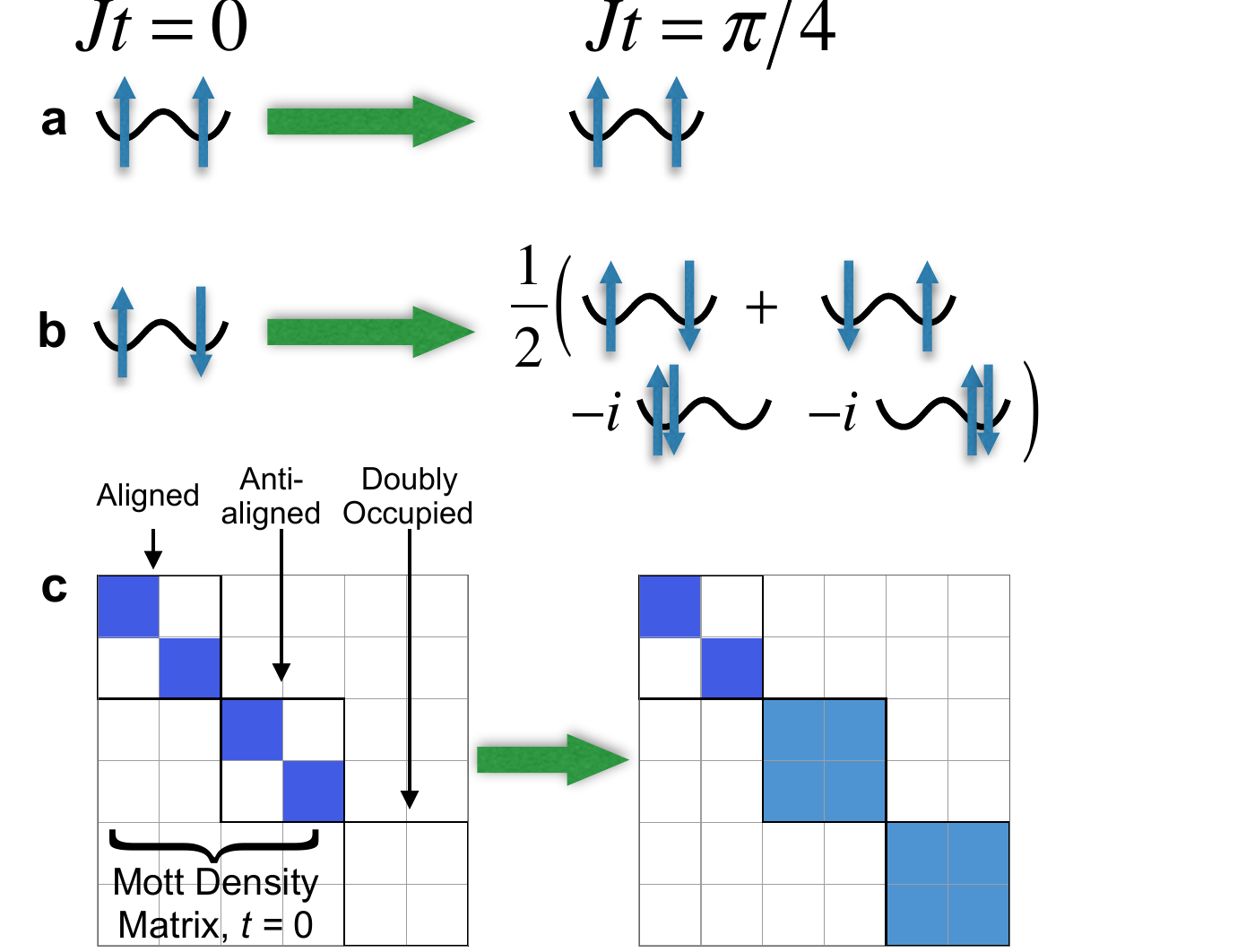}
\caption{\label{fig: 2 sites}Schematic diagram of the time evolution of a two-site model at unit filling, initially with $J\ll T \ll U$, and then quenched to $U=0$. (a) Aligned initial states do not evolve. (b) Anti-aligned initial states (the two-site equivalent of AFM initial states) evolve into superpositions with weight on doubly-occupied states at later times. (c) The Mott insulator-like initial density matrix on two sites transfers some weight from its matrix elements for anti-aligned states to doubly-occupied states at later times, while aligned states' matrix elements do not change. The matrix elements' magnitudes are indicated by color, from white (zero) to  dark blue (maximal).}
\end{figure}

To qualitatively understand the correlation dynamics, it is useful to consider the dynamics of a two-site model, which is shown schematically in Fig. \ref{fig: 2 sites}. Let $\ket{p\,q}$ denote a state with $p$ and $q$ referring to the left and right sites respectively, and taking on the values $0$ (empty), $\uparrow$ (one atom with spin up), $\downarrow$ (one atom with spin down), and $d$ (two atoms).
The state $\ket{\uparrow \, \uparrow}$ does not evolve since Pauli blocking prevents it from coupling to any other states, while the state $\ket{\uparrow \, \downarrow}$ evolves in the Schr{\"o}dinger picture as
\begin{multline}
\ket{\uparrow \, \downarrow}(t)=\cos^{2}(t)\ket{\uparrow \, \downarrow}_{0}+\sin^{2}(t)\ket{\downarrow \, \uparrow}_{0}\\{}-i\cos(t)\sin(t)\left(\ket{d \, 0}_{0}+\ket{0 \, d}_{0}\right)
\end{multline}
as shown in Figs. \ref{fig: 2 sites}(a) and \ref{fig: 2 sites}(b) for $J t = \frac{\pi}{4}$. 
In Fig. \ref{fig: 2 sites}(c), the ferromagnetic character of the dynamic spin correlations becomes apparent by observing that although the initial density matrix has equal weight on aligned (e.g. $\ket{\uparrow \, \uparrow}$) and anti-aligned (e.g. $\ket{\uparrow \, \downarrow}$) spin configurations, the time-evolved matrix has more weight on the aligned states. 
This is because the aligned states stay frozen in time, while the anti-aligned states can partially convert to states with doublons and holes, reducing their spin correlations.
The density correlations can be explained similarly: the initial density matrix has no weight on doubly-occupied states, but the time-evolved matrix does.
The double-occupancy next to a vacant site represents a negative two-site density correlation, or equivalently, a (short-ranged) density wave correlation.

Local observables approach constant values at large times. 
Although the noninteracting system is clearly integrable and thus not expected to thermalize, the expectation values of the local observables as $t \rightarrow \infty$ are consistent with those of a thermal equilibrium state, in particular one at $T = \infty$. 
This occurs because the initial state is a product state in the site basis. Thermalization is not expected for other, more general initial states.

\section{\label{sec:compare}Quenches from more general initial states: doped and spin-imbalanced systems, 2-dimensional Mott insulators, and antiferromagnets}

\begin{figure}
\includegraphics[clip=true, trim= 0mm 0mm 170mm 120mm, width= 4.00in, keepaspectratio=true]{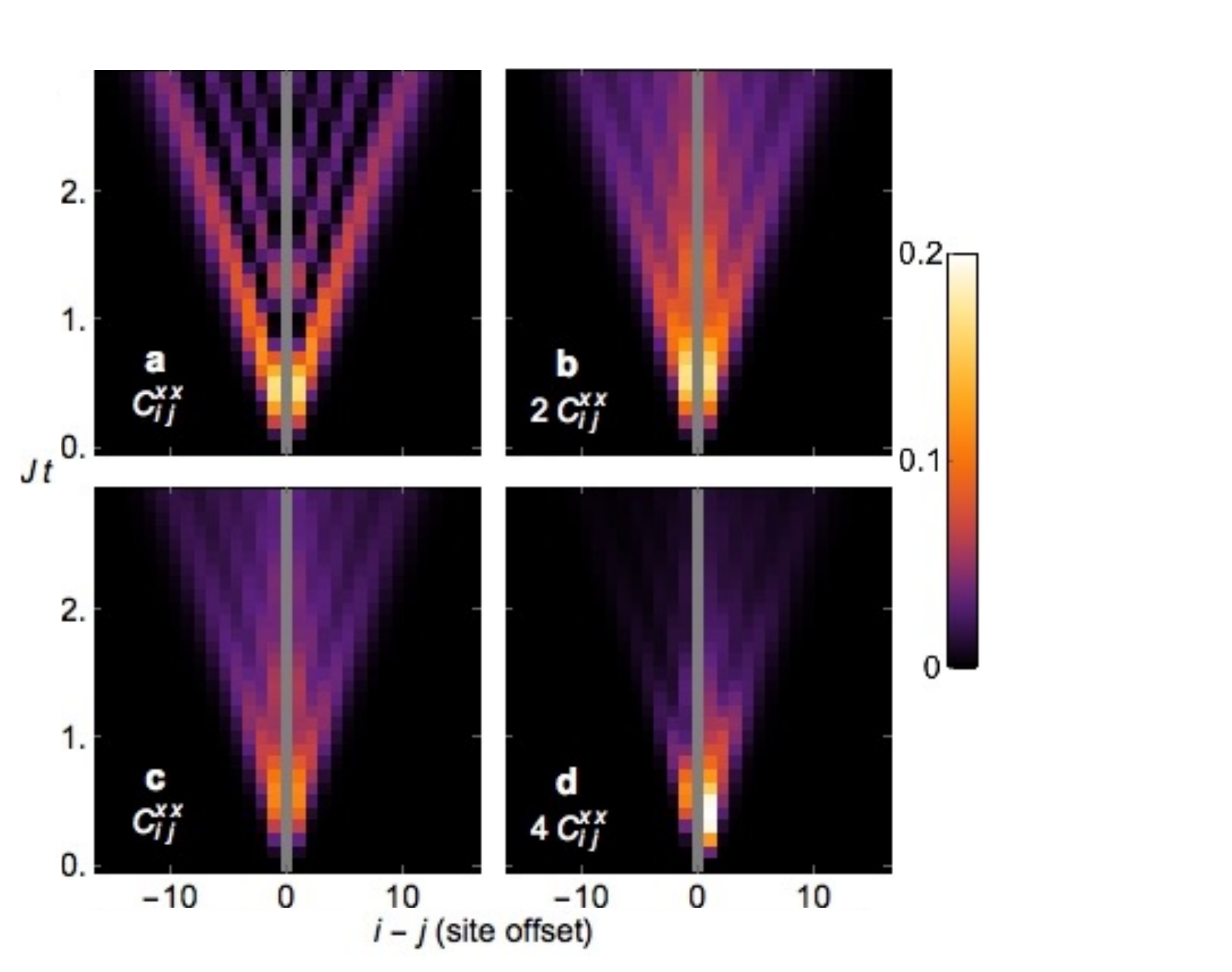}
\caption{\label{fig: comparison}Spreading of correlations is generic, demonstrated by four additional classes of initial conditions.  (a) A 1D product state antiferromagnet aligned along the $z$-axis at half-filling, (b) a $J\ll T \ll U$ 1D hole-doped system, with $ \langle n \rangle \approx 0.85$, (c) a $J\ll T \ll U$ spin-imbalanced 1D system with $ \langle \sigma^z \rangle \approx 0.25$, and (d) a $J\ll T \ll U$ 2D Mott insulator. Site offsets to the right are along the (1,0) direction, and those to the left are along (1,1).}
\end{figure}

The light-cone spreading of correlations from an uncorrelated initial state is not restricted to a 1D Mott insulator, but also occurs for a variety of initial conditions, as shown in Fig. \ref{fig: comparison}. 
 We demonstrate this for a $T\gg J$ metal (with $n<1$), a spin imbalanced $T\gg J$ Mott insulator, a product state antiferromagnet, and a 2D $T\gg J$ Mott insulator. We note that the metal can be alternatively viewed as a doped Mott insulator when $U\gg J$.
 
Although both the spin imbalanced and hole-doped 1D initial states show correlations developing in light cones as in the Mott insulator, the magnitude of the correlations is reduced, as shown in Fig. \ref{fig: comparison}; this follows from Eq.~\eqref{eq: factored time evolution eq} with the $f$ and $g$ in Eq.~\eqref{eq: f and g} evaluated in these limits (see below). 
One can induce a partially spin-polarized initial state by adding a term $B S^z_i$ to Eq.~(\ref{eq:single-site H}), and likewise induce a number density other than one per site by taking the chemical potential to be $\mu=U/2+\Delta$ with $\Delta \ne 0$.

For a partially spin-polarized system at unit filling, one finds
\begin{subequations}
\label{eq: f and g FM}
\begin{eqnarray}
 \hspace{-1cm}f_{\alpha\beta}^{i} &=&
	\frac{1}{\mathcal{N}_{1}}\delta_{\alpha\beta}\left(1+e^{\frac{1}{2}\tilde{U}+\frac{1}{2}\sigma^{z}_{\alpha\beta}\tilde{B}}\right) \label{eq: f FM}\\
\hspace{-1cm}g_{\alpha\beta\gamma\delta}^{i} &=& \begin{cases}
	f_{\alpha\alpha}^{i} &\alpha=\beta=\gamma=\delta\\
	\frac{1}{\mathcal{N}_{1}} &\alpha=\beta\ne\gamma=\delta\\
	\frac{1}{\mathcal{N}_{1}}\left(1+e^{\frac{1}{2}\tilde{U}+\frac{1}{2}\sigma^{z}_{\alpha\delta}\tilde{B}}\right)&\alpha=\delta\ne\beta=\gamma\\
	0 &\rm{otherwise}\\
\end{cases} \label{eq: g FM}
\end{eqnarray}
\end{subequations}
with the normalization factor 
\begin{equation}
\mathcal{N}_{1}=2+2\exp\left(\frac{1}{2}\tilde{U}\right)\cosh\left(\frac{1}{2}\tilde{B}\right).
\end{equation}
For a system doped away from unit filling,
\begin{subequations}
\label{eq: f and g doped}
\begin{eqnarray}
f_{\alpha\beta}^{i} &=& 
	\frac{1}{\mathcal{N}_{2}}\delta_{\alpha\beta}\left(e^{\tilde{\Delta}}+e^{\frac{1}{2}\tilde{U}}\right) \label{eq: f doped}\\
g_{\alpha\beta\gamma\delta}^{i} &=& \begin{cases}
	f_{\alpha\alpha}^{i} &\alpha=\beta=\gamma=\delta\\
	\frac{1}{\mathcal{N}_{2}}e^{\tilde{\Delta}} &\alpha=\beta\ne\gamma=\delta\\
	\frac{1}{\mathcal{N}_{2}}e^{\frac{1}{2}\tilde{U}} &\alpha=\delta\ne\beta=\gamma\\
	0 &\rm{otherwise}\\
\end{cases} \label{eq: g doped}
\end{eqnarray}
\end{subequations}
with 
\begin{equation}
\mathcal{N}_{2}=2\left[\cosh\left(\tilde{\Delta}\right)+\exp\left(\frac{1}{2}\tilde{U}\right)\right].
\end{equation}

Finally, for dynamics initiated from a 1D antiferromagnetic product state given by 
\begin{equation} \rho=\bigotimes_{i}\begin{cases}\ket{\uparrow}_{i}\bra{\uparrow}_{i} & i \quad\rm{even}\\ \ket{\downarrow}_{i}\bra{\downarrow}_{i} & i \quad\rm{odd}\end{cases}
\end{equation}
one finds
\begin{subequations}
\label{eq: f and g AFM}
\begin{eqnarray}
f_{\alpha\beta}^{i} &=& \begin{cases}
	\delta_{\alpha\beta}\delta_{\alpha\uparrow}& \quad i \quad\rm{ even}\\
	\delta_{\alpha\beta}\delta_{\alpha\downarrow}& \quad i \quad\rm{odd}\\
\end{cases}\label{eq: f AFM}\\
g_{\alpha\beta\gamma\delta}^{i} &=& f_{\alpha\delta}^{i}\delta_{\beta\gamma} \label{eq: g AFM}
\end{eqnarray}
\end{subequations}

\begin{figure}
\includegraphics[clip=true, trim= 0mm 0mm 180mm 120mm,width= 4.00in, keepaspectratio=true]{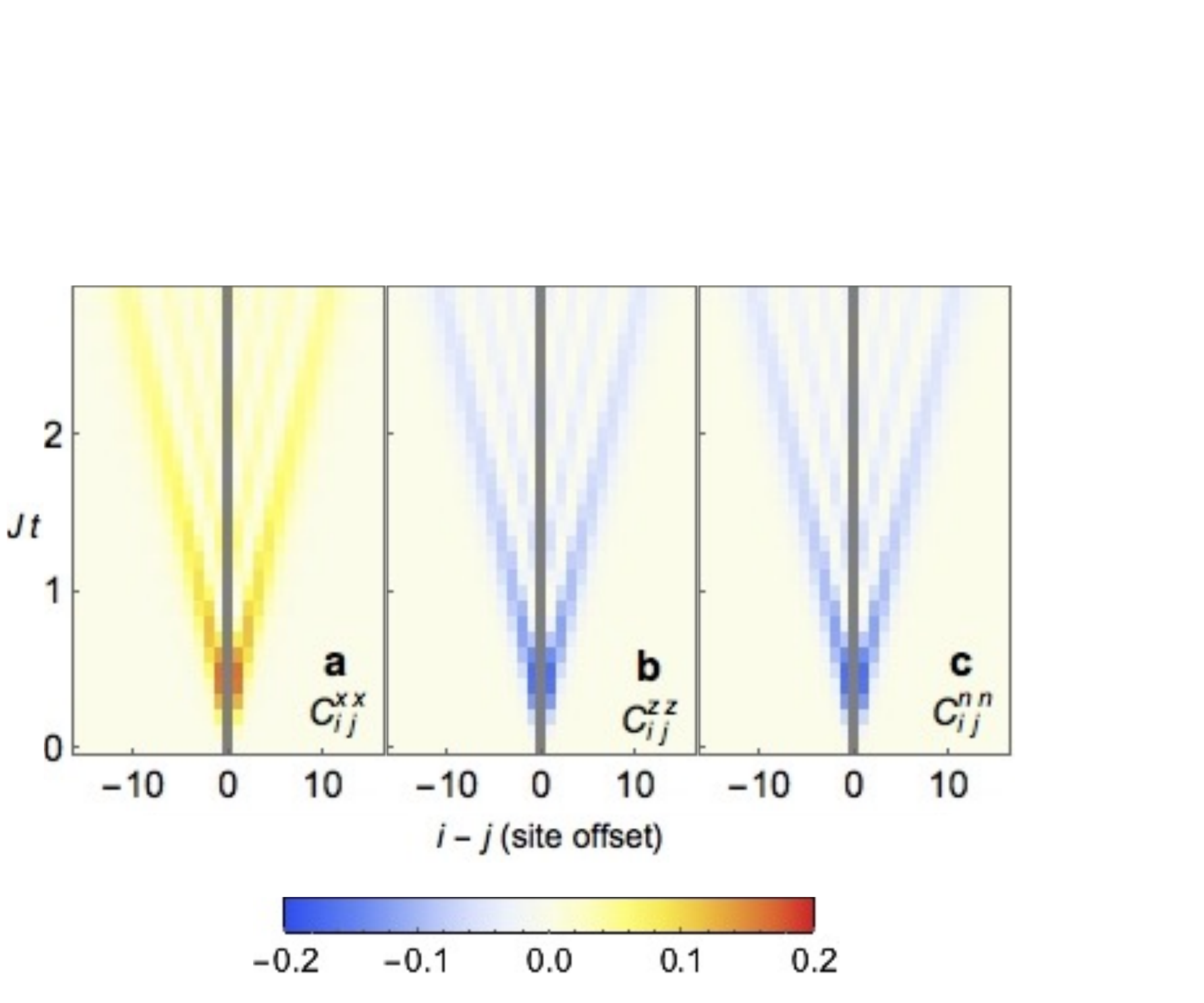}
\caption{\label{fig:AFM}Spreading of correlations in a 1D product state antiferromagnet aligned along the $z$-axis. (a) Spin-spin correlations in the $x$-direction $C^{xx}_{ij}$. The $y$-$y$ correlations are identical: $C^{yy}_{ij}=C^{xx}_{ij}$ (b) Spin-spin correlations in the $z$-direction $C^{zz}_{ij}$. (c) Density-density correlations $C^{nn}_{ij}$.}
\end{figure}

The antiferromagnet-initiated dynamics displays a distinctive feature: anisotropy in the spin correlations. 
As shown in Fig. \ref{fig:AFM}, the $C^{xx}$ and $C^{yy}$ connected correlations remain positive and equal in magnitude, as they were in previous cases, but the $C^{zz}$ and $C^{nn}$ connected correlations are negative. They are, however, still equal in magnitude. The magnitude of the correlations is larger than those of the 1D Mott insulator.
The anisotropy manifests despite the SU(2) symmetry of the Hamiltonian due to the broken symmetry of the initial state.

The light-cone spreading of correlations is not restricted to one-dimensional systems. 
As seen in Fig. \ref{fig: comparison}(d) a two-dimensional Mott insulator on a square lattice shows qualitatively similar dynamics, but develops weaker transient correlations than a 1D Mott insulator with the same post-quench tunneling amplitude $J$. 
 
In a two-dimensional Mott insulator, the initial expectation values do not differ from the 1D case, but the propagators take a different form. For a square lattice, they are 
\begin{equation}
A_{\bm{p}\bm{q}}(t)=A_{p_{x}q_{x}}(t)A_{p_{y}q_{y}}(t)\label{eq: 2D prop}%\left(-i\right)^{\left\Vert\bm{p}-\bm{q}\right\Vert_{1}}\mathcal{J}_{\left|p_{x}-q_{x}\right|}\left(2Jt\right)\mathcal{J}_{\left|p_{y}-q_{y}\right|}\left(2Jt\right) 
\end{equation}
where $\bm{p}$ and $\bm{q}$ are integer vectors indicating sites on the square lattice.
Note that the 2D propagators factor  into 1D components in this way due to the properties of the square lattice.
It could be interesting to explore the effects of other geometries, where interference between different paths can give propagators with structures other than Eq.~\eqref{eq: 2D prop}.

\section{\label{sec:hole transport}Hole transport dynamics}

\begin{figure*}
\includegraphics[clip=true, trim= 10mm 0mm 25mm 50mm, width= 8in, keepaspectratio=true]{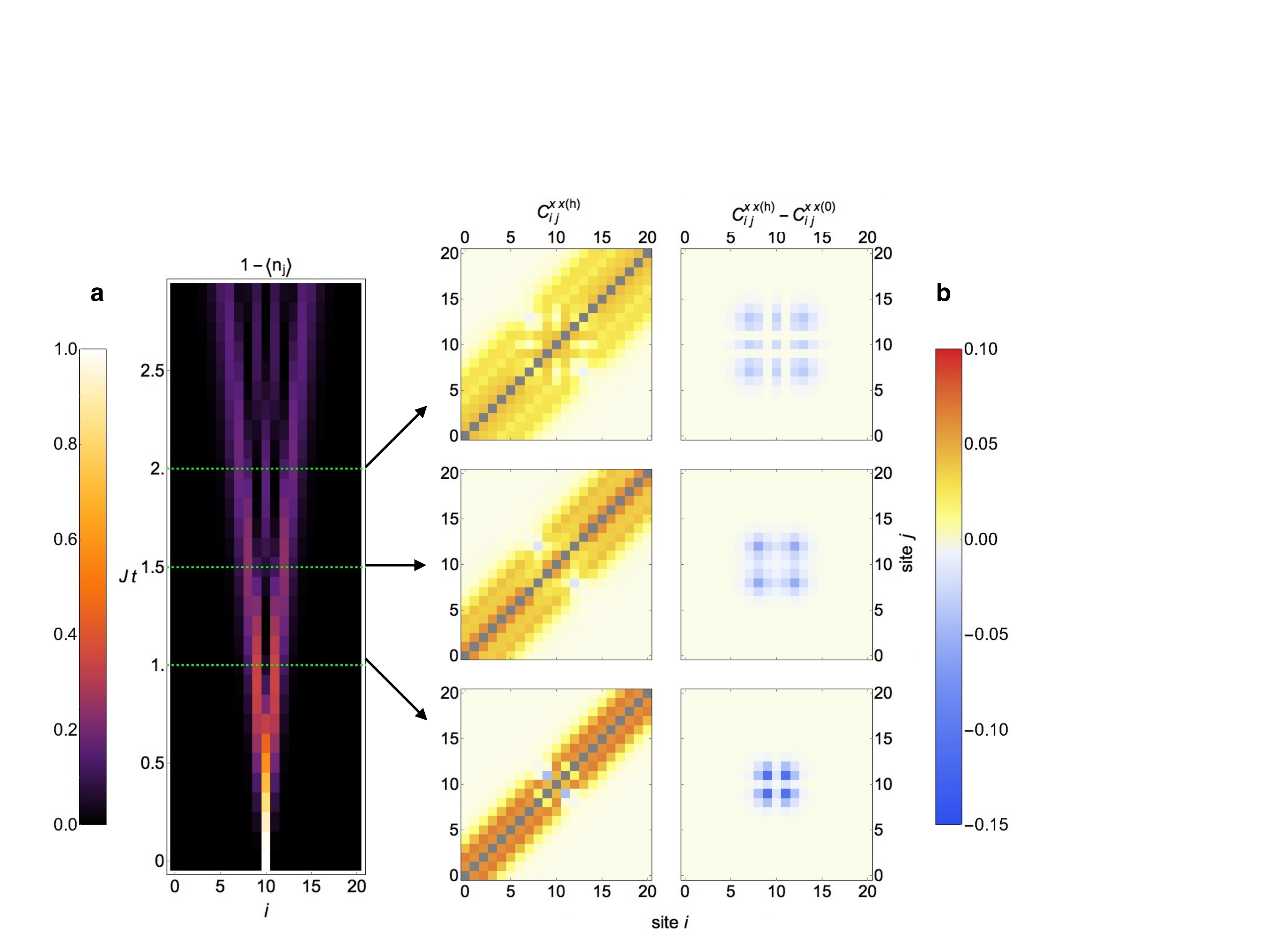}
\caption{\label{fig:hole}Dynamics of a hole initially at $j_0=10$ on a $J\ll T \ll U$ Mott insulating background after quenching to $U=0$, and the hole's influence on spin correlations. (a) Hole density $1-\langle n_{j} \rangle$. (b) Connected spin correlations $C^{xx}_{ij}$ (left) and the difference in connected correlations between this system and a uniform system, with no hole (right). These are shown at the times $Jt = 1$ (bottom), $Jt = 1.5$ (middle), and $Jt = 2$ (top). The center of each plot in (b) corresponds to $i=j=10$.}
\end{figure*}

Now we consider a system in which a single hole is added, localized to a single site, to the $T\gg J$ Mott insulator state discussed in Section \ref{sec:MI}. 
The behavior of hole defects in fermionic systems underpins the physics of many strongly correlated materials, where doping Mott insulators leads to a panoply of intriguing phenomena, most famously high-temperature superconductivity.
The topic has long been of interest
\cite{Montambaux1982,PhysRevB.41.892},
and thermodynamics, spectral properties, and dynamics have been investigated
\cite{PhysRevA.74.020102,PhysRevB.90.024404,PhysRevB.90.205126,PhysRevB.90.035145}.
It has recently been shown that a hole defect in a Mott insulating system disperses neither purely ballistically nor diffusively: the hole in fact leaves a trace in the background as it travels, preventing the quantum interference of some trajectories \cite{PhysRevLett.116.247202}. 
In light of this interesting result it is useful to consider the noninteracting analog.

The initial density matrix for a hole initially at site $j_0$ in a $T\gg J$ Mott insulator is 
\begin{equation}
\rho=Z^{-1}\bigotimes_{i}\begin{cases}\ket{0}_{i}\bra{0}_{i} & i = j_0\\ \exp\left(-\beta H_{i}\right) &  \rm{otherwise}\end{cases} 
\end{equation}
which is identical to the Mott insulator density matrix except at $j_0$. Likewise $f^{j_0}$ and $g^{j_0}$ are zero, with $f$ and $g$ otherwise identical to those of the Mott insulator.

We find that, as expected for single particle ballistic motion, the hole disperses outwards according to the distribution 
\begin{equation}
1-\left\langle n_{j}\right\rangle (t)=\left|\mathcal{J}_{\left|j-j_{0}\right|}\left(2Jt\right)\right|^{2}
\end{equation}
as shown in Fig. \ref{fig:hole}(a).

Figure \ref{fig:hole}(b) shows that as the hole disperses, it modifies spin correlations between pairs of nearby sites. 
In particular, the correlations obtain contributions of the opposite sign, suppressing the correlations and even at some points reversing their sign, compared to their values in the absence of the hole. 
This gives the impression that the spreading hole is dressed with a cloud of spin correlations. 
Such a phenomenon might be thought to be unique to an interacting system, and is certainly of interest there. 
Our results show that apparently similar phenomena occur even without interactions, although they arise  from different causes.

\section{\label{sec:conclusion}Conclusions}

We have shown that interaction quenches of the Fermi-Hubbard model from initial product states to the non-interacting limit produce transient connected two-site correlations. 
The correlations develop despite the initial states being at high temperature or, in the case of the product state antiferromagnet, high energy.
Even when the temperature is much greater than the initial tunneling, and very much greater than the superexchange energy scale, significant correlations exist in the dynamics.

This finding contrasts with the natural idea that at high temperature or high entropy correlations should be absent. 
For example, in the context of previous work that has observed correlations out of equilibrium from a high-entropy initial state of ultracold molecules
~\cite{nature-501,PhysRevLett.111.185305, PhysRevLett.113.195302} 
it has sometimes been argued that long-range interactions are crucial.
Our present work shows that in contrast, correlations are ubiquitous out of equilibrium, even when one starts from high entropy states.

We generally observe that the correlations grow in a light cone and then fade away. 
In the process, interesting structures emerge, such as correlations with a sign opposite that of the equilibrium system and intertwined density and spin correlations of equal magnitude.
It is noteworthy that these and other phenomenon, such as the appearance of a spin correlation cloud around a hole, that look intriguing and might typically be associated with strong interactions, can occur quite generally in out-of-equilibrium systems, even in the absence of interactions. 
In this light, our work provides a useful comparison for future work with interacting quenches.

We note that the peak magnitudes of the connected correlations are up to about $\sim0.15$. 
While the precise values depend strongly on the initial conditions, these values are comparable in magnitude to recent equilibrium observations of correlations in 1D~\cite{Boll1257} and 2D~\cite{Parsons1253}, and 3D~\cite{PhysRevLett.106.030401,PhysRevLett.107.086401,AFM-corr-Hulet}.
This indicates that the correlations generated dynamically from uncorrelated initial states are large enough to be experimentally measured.

One interesting future direction involves understanding how the features of integrability of this system manifest for finite temperature initial states. 
This is much less explored than quenches from low temperature initial states. It is expected to be fruitful to start by understanding the simplest integrable systems -- non-interacting ones.
In our system, even though one expects the steady state to be non-thermal, i.e. to prethermalize, due to our initial conditions the prethermalization coincides with a $T=\infty$ thermal equilibrium state as measured by spin, density, and correlation operators. 
This is because the initial product state has equal overlap with all the states in the noninteracting band (i.e., the eigenstates of the final Hamiltonian), leading to a final state with occupation numbers independent of energy.
We expect that perturbing the initial state away from a perfect product state will lead to a detectable difference from a thermal steady state. 
Likewise, including weak interactions in the post-quench Hamiltonian should allow for the investigation of integrability breaking and prethermalization.

\acknowledgements 

We thank Rafael Poliseli-Teles for conversations. 
K.R.A.H. thanks the Aspen Center for Physics, which is supported by National Science Foundation grant PHY-1066293, for its hospitality while part of this work was carried out, and the Welch foundation, Grant No. C-1872.
R.G.H. thanks the National Science Foundation (Grant No. PHY-1408309), the Welch Foundation (Grant No. C-1133), an ARO-MURI (Grant No. W911NF-14-1-0003), and the ONR.

%\begin{figure}
%\includegraphics[clip=true, trim= 12mm 0mm 170mm 100mm, width= 4.00in, keepaspectratio=true]{FH_Paper_Figures/Figure3.pdf}
%\caption{\label{fig:wide}Spin-spin correlations as a function of site-offset and time, for four different systems, all initially in the $U \gg J,T$ regime. (a) A 1D antiferromagnet at half-filling, (b) a hole-doped 1D system, with n ~ 0.85, (c) a spin-imbalanced  1D system with Sz ~ 0.25, and (d) a two-dimensional Mott insulator. In (d) site offsets to the right are in the (10) direction, and those to the left are in the (11) direction.}
%\end{figure}

\bibliography{citations,citations-continued,citations-3,citations-4,citations-5}{}

\end{document}